\documentclass[conference,compsoc]{IEEEtran}
\pdfoutput=1
\IEEEoverridecommandlockouts
\usepackage{cite}
\usepackage{amsmath,amssymb,amsfonts}
\usepackage{graphicx}
\usepackage{textcomp}
\usepackage{xcolor}
\usepackage{url}
\usepackage{hyperref}
\hypersetup{
    colorlinks=true,
    linkcolor=blue,
    filecolor=magenta,      
    urlcolor=cyan,
    pdfpagemode=FullScreen,
    }
\usepackage{listings}
\usepackage{cuted}
\usepackage{algorithm}
\usepackage{algpseudocode}
\usepackage{caption}
\usepackage{subcaption}
\usepackage{multirow}
\usepackage{array}

\def\BibTeX{{\rm B\kern-.05em{\sc i\kern-.025em b}\kern-.08em
    T\kern-.1667em\lower.7ex\hbox{E}\kern-.125emX}}
\begin{document}

\title{\textbf{Enhancing Adversarial Example Detection Through Model Explanation}}

\author{\IEEEauthorblockN{Qian Ma\IEEEauthorrefmark{1}}
\IEEEauthorblockA{\textit{College of Information Sciences and Technology}\\
\textit{The Pennsylvania State University}\\
\rm State College, USA \\
masonma@psu.edu}\\
\and
\IEEEauthorblockN{Ziping Ye\IEEEauthorrefmark{1}}
\IEEEauthorblockA{\textit{Department of Computer and Information Science} \\
\textit{University of Pennsylvania}\\
\rm Philadelphia, USA\\
zipingy@seas.upenn.edu}

\thanks{\IEEEauthorrefmark{1}Qian Ma and Ziping Ye contributed equally to this research.}
}

\maketitle

\begin{abstract}
    Adversarial examples are a major problem for machine learning models, leading to a continuous search for effective defenses. One promising direction is to leverage model explanations to better understand and defend against these attacks. We looked at AmI, a method proposed by a NeurIPS 2018 spotlight paper that uses model explanations to detect adversarial examples. Our study shows that while AmI is a promising idea, its performance is too dependent on specific settings (e.g., hyperparameter) and external factors such as the operating system and the deep learning framework used, and such drawbacks limit AmI's practical usage. Our findings highlight the need for more robust defense mechanisms that are effective under various conditions. In addition, we advocate for a comprehensive evaluation framework for defense techniques.
\end{abstract}


\section{Introduction}
Adversarial example \cite{goodfellow2014explaining, moosavi2016deepfool, carlini2017towards, papernot2016limitations, kurakin2016adversarial, liu2019transferable, kurakin2016adversarialscale, ilyas2018black, zhao2017adversarial, Szegedy2014Intriguing} is a long-standing and widespread problem in machine learning, making models vulnerable to various attacks. Despite the continuous advances in defense strategies \cite{papernot2016distillation, madry2017towards, xu2018feature, cohen2019certified, meng2017magnet, song2017pixeldefend, lecuyer2019certified, samangouei2018defense, zheng2016improving, wong2018provable}, sophisticated and innovative attacks quickly defeat such efforts. Adversarial examples often differ from clean examples by only minor variations. For example, in the image domain, the differences between adversarial examples generated by various algorithms and benign examples are imperceptible to the human eye. On the other hand, explainable machine learning methods \cite{lime, shap} aim to improve human understanding of how models make predictions and identify the features that are most significant. This naturally suggests a defense strategy using model explanations.

An explanation-assisted adversarial example detection method, "Attacks Meet Interpretability" (AmI) \cite{AmI}, was introduced in a spotlight paper at NeurIPS 2018. This method involves identifying and adjusting the activation of neurons corresponding to key image features to distinguish between adversarial and clean examples. However, subsequent analyzes, such as \cite{NC}, have raised concerns regarding AmI \cite{AmI}'s robustness to certain types of attacks, underscoring the need for further investigation.

\begin{figure}[htbp]
\centerline{\includegraphics[width=0.47\textwidth]{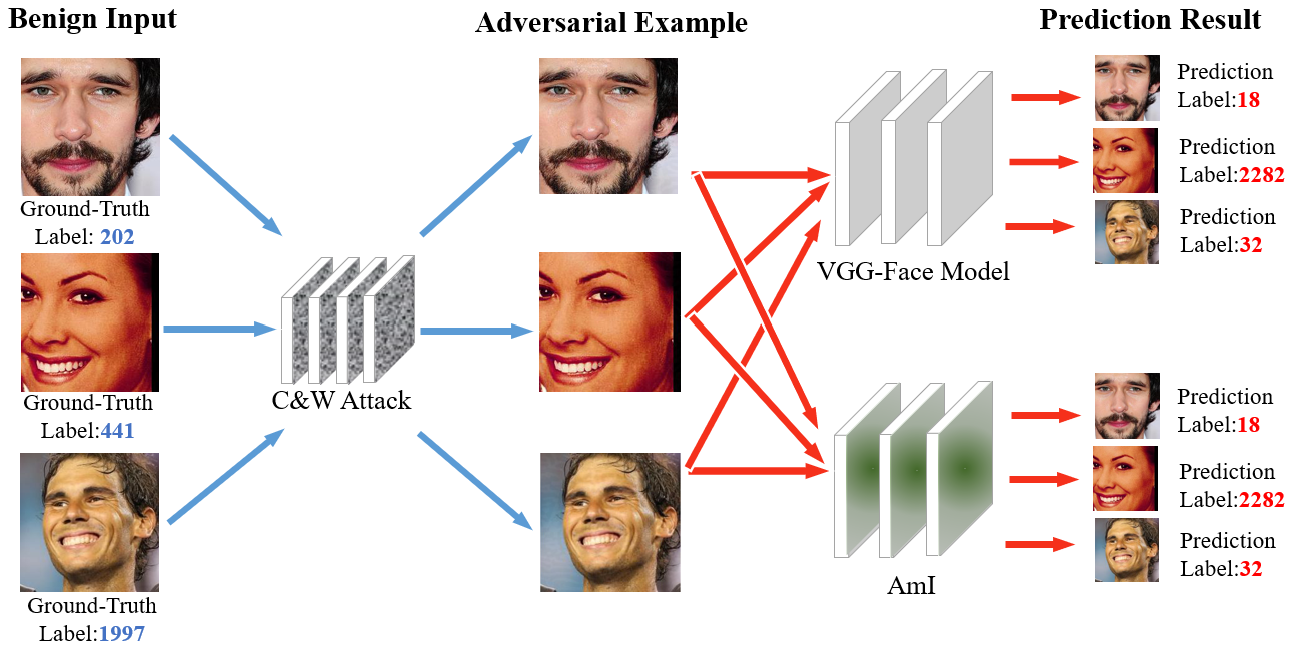}}
\caption{AmI \cite{AmI} is unable to detect these adversarial examples since their predictions are identical to the VGG model \cite{frs} result.}
\label{fig:x-y-y}
\end{figure}

Inspired by \cite{NC} and given the novelty and potential of AmI \cite{AmI}, our study delves into a comprehensive examination of its capabilities and limitations. We found that AmI \cite{AmI}'s effectiveness is significantly affected by both its hyperparameter configurations and external variables, such as the operating system and the deep learning framework in use. Furthermore, we pinpoint a flaw in AmI \cite{AmI}'s evaluation methodology: the test set only contains adversarial examples, which inherently biases the evaluation of its detection accuracy by overlooking the potential for false positives.

In summary, our study makes the following contributions:
\begin{itemize}
    \item We conduct a comprehensive evaluation of AmI \cite{AmI}, both conceptually and experimentally, by encompassing a broader spectrum of scenarios than previously considered.
    \item We find the sensitivity of AmI \cite{AmI} to both internal settings and external factors regarding its robustness as a detection technique.
    \item We advocate for a more rigorous evaluation framework for defense techniques that considers both false positives and false negatives, ensures robustness against various factors, and emphasizes practical applicability.
\end{itemize}

The rest of this paper is structured as follows: we begin with an overview of AmI \cite{AmI} and adversarial attacks. Subsequently, we present our empirical study and observations, and experimental results. We then examine the broader implications of our findings and suggest future research directions.

Our code can be found at \href{https://anonymous.4open.science/r/Enhancing-Adversarial-Example-Detection-Through-Model-Explanation-0B6F}{this link}.



\section{Preliminaries}
\subsection{Attacks Meet Interpretability (AmI)}
NeurIPS 2018 spotlight paper \cite{AmI} proposes an explanation-assisted adversarial example detection method, called Attacks Meet Interpretability (AmI). This approach establishes a novel bidirectional link between features and internal neurons. Specifically, AmI \cite{AmI} enhances the activation values of neurons associated with features understandable by humans (referred to as witness neurons) while reducing the activation of other neurons (non-witness neurons). The underlying premise is that accurate predictions should rely on features that are understandable to humans. In the realm of face recognition, such features would include elements such as the eyes, nose, and mouth. AmI \cite{AmI} distinguishes adversarial examples by comparing the model predictions before and after the adjustment of neuron activation. A discrepancy in the predictions indicates that the original model prediction (based on the VGG face model \cite{frs}) did not depend on human-interpretable features, leading to detection of an adversarial example.

We briefly describe the three key steps in AmI \cite{AmI}. First, it recognizes neurons that are associated with individual attributes (witnesses) and neurons that are not necessarily essential to the attribute witnesses (non-witnesses) in the original model. Then, it extracts the witnesses and non-witnesses neurons from the original model. Lastly, it leverages the extracted witnesses to construct a new attribute-steered model by strengthening the witness neurons and weakening the non-witnesses neurons. AmI \cite{AmI} determines the presence of adversarial examples based on discrepancies in prediction outcomes.

\subsection{Adversarial Attacks}
The C\&W attack \cite{wb8}, recognized as one of the most effective methods for creating adversarial examples, operates within the image domain under the assumption of complete white-box access to the neural network, including its architecture and parameters. Due to the transferability of adversarial examples \cite{transferability}, it is possible to train a substitute model with only black-box access to the target model. The adversarial examples crafted using the substitute model are highly likely to be effective against the target model.

Given the result that AmI \cite{AmI} claims it can gain at least $90\%$ detection accuracy against C\&W attack  \cite{wb8}, \cite{NC} expresses concerns regarding AmI \cite{AmI}’s robustness to C\&W attack \cite{wb8}.



\section{Empirical Study and Observation}

\subsection{Neuron Weakening and Strengthening}
AmI \cite{AmI} employs three hyperparameters to precisely modulate the activation values of neurons, either by weakening or strengthening them. This approach enables the determination of the extent to which features and their corresponding neurons are interconnected by adjusting these parameters.

\textbf{Neuron weakening hyperparameter $\boldsymbol{\alpha}$.} $\alpha$ defines the magnitude of weakening and is capable of reducing the values of non-witness neurons in each layer that have activation values larger than the mean values of witness neurons in that layer. The following reduction is performed on all the non-witness neurons. \textit{v} denotes the value of a non-witness neuron, $\mu$ and $\sigma$ are the mean and standard deviation of values of witness neurons respectively \cite{AmI}. The default $\alpha$ is set to $100$.

\begin{align*}
        v^{'} = e^{-\frac{v-\mu}{\alpha \cdot \sigma } } \cdot v
\end{align*}

\textbf{Neuron strengthening hyperparameter $\boldsymbol{\beta}$.} $\beta$ is not defined in \cite{AmI}. From the context, we believe that $\beta$ represents the magnitude of the strengthening. The default value $\beta$ is set to $60$.

\textbf{Strengthening factor hyperparameter $\boldsymbol{\epsilon}$.} $\epsilon$ is the strengthening factor and the default value is $1.15$. $\beta$ and $\epsilon$ are capable of expanding the values of all witness neurons as follows.
\begin{align*}
        v^{'} = \epsilon  \cdot v  +  (1 - e^{-\frac{v-min}{\beta \cdot \sigma } }) \cdot v
\end{align*}

The corresponding code is as follows.
\definecolor{CPPViolet} {HTML} {7040A0}
\lstset{
    columns=fixed,       
    frame=none,                                         
    backgroundcolor=\color[RGB]{245,245,244},           
    keywordstyle=\color[RGB]{40,40,255},                 
    language=python,                                        
    morekeywords={alignas,continute,friend,register,true,alignof,decltype,goto,
    reinterpret_cast,try,asm,defult,if,return,typedef,auto,delete,inline,short,
    typeid,bool,do,int,signed,typename,break,double,long,sizeof,union,case,
    dynamic_cast,mutable,static,unsigned,catch,else,namespace,static_assert,using,
    char,enum,new,static_cast,virtual,char16_t,char32_t,explict,noexcept,struct,
    void,export,nullptr,switch,volatile,class,extern,operator,template,wchar_t,
    const,false,private,this,while,constexpr,float,protected,thread_local,
    const_cast,for,public,throw,std},
    emph={map,set,multimap,multiset,unordered_map,unordered_set,
    unordered_multiset,unordered_multimap,vector,string,list,deque,
    array,stack,forwared_list,iostream,memory,shared_ptr,unique_ptr,
    random,bitset,ostream,istream,cout,cin,endl,move,default_random_engine,
    uniform_int_distribution,iterator,algorithm,functional,bing,numeric,},
    emphstyle=\color{CPPViolet}, 
}              
\begin{lstlisting}
def strengthen(x):
    return 2.15 - np.exp(-x/60)
\end{lstlisting}

From our preliminary experiments, we observed that the prediction accuracy of AmI \cite{AmI}, as reported, could not be reproduced using the default $\beta$ value of 60. We observed a significant fluctuation in the prediction accuracy rate with variations in $\beta$. However, \cite{AmI} does not provide a clear definition of $\beta$, nor does it disclose that the claimed high detection accuracy is achieved by adjusting $\beta$ from its default setting.

\begin{table}[h]
\centering
\caption{Conceptual Analysis of AmI.}
\begin{tabular}{ |m{1.2cm} | m{1.7cm}| m{3cm}| m{.9cm}| } 
 \hline
 Case & Meaning & Explanation & Does AmI detect it? \\ 
 \hline \hline
 X - X - X & True Negative & Benign example. The original model is not fooled. & No \\ 
 \hline
  X - X - Y & False Positive & AmI mark the benign example as adversarial. & Yes \\ 
 \hline
 X - Y - X & True Positive & Original model is fooled, but the attribute model is not. AmI is robust to this adversarial example. & Yes \\ 
 \hline
 X - Y - Y & False Negative & Original model and attribute model are fooled in the same way (i.e. misclassified in the same way). \cite{NC} uses these samples to evaluate AmI. & No \\  
 \hline
 X - Y - Z & True Positive & Both the original model and the attribute model are fooled, but differently. & Yes \\  
 \hline
\end{tabular}
\label{tablexyz}
\end{table}

\subsection{Conceptual Analysis}

To evaluate AmI \cite{AmI}, we reference the five cases detailed in Table \ref{tablexyz}, involving three critical values: the ground truth (X), the prediction by the original model (Y), and attribute model's prediction (Z), represented in the format X - Y - Z.

Two pairs of values deserve particular attention for this assessment. When the ground truth (X) aligns with the original model prediction (Y), the instance is benign. In contrast, a discrepancy indicates an adversarial example. The comparison between Y and Z dictates AmI's assessment: identical values indicate that an example is not identified as adversarial, while different values prompt AmI to classify it as adversarial. This comparison facilitates the identification of true positives, false positives, true negatives, and false negatives, adhering to standard definitions in evaluation metrics.

\section{Experimental Results}
\subsection{Robustness to Hyperparameter Changes}

\textbf{The default $\boldsymbol{\beta}$.} Using the patch attack \cite{patch} as an example, we conducted the "patch first" and "patch next" attacks against AmI \cite{AmI} with patching examples. Our findings, which show accuracy rates of 0.19 and 0.40 respectively (as detailed in Table \ref{results by tuning beta}), diverge significantly from those reported in \cite{AmI}. This inconsistency is observed across different types of attacks as shown in Table \ref{results by tuning beta}. Here, 'accuracy' refers to the portion of examples correctly identified as adversarial out of the total number of adversarial examples used in the attack. Despite using the same code and dataset within the Caffe framework to reproduce the results, we were unable to derive the same results claimed by \cite{AmI}.

\begin{table}[h]
\centering 
\caption{Achieve the same detection accuracy as AmI \cite{AmI} presents by tuning $\boldsymbol{\beta}$ in various attacks.} 
\label{results by tuning beta}
\renewcommand\arraystretch{1.5}
\setlength{\tabcolsep}{1mm}{
\begin{tabular}{cllcccccc}
\hline
                                                                                            &  &  & \multicolumn{4}{c}{Targeted}                                                                                                                                                                                                                          & \multicolumn{2}{c}{Untargeted}                                                                                            \\ \cline{4-9} 
                                                                                            &  &  & \multicolumn{2}{c}{Patch}                                                                                               & \multicolumn{2}{c}{Glasses}                                                                                                 &                                                             &                                                             \\ \cline{4-7}
\multirow{-3}{*}{AmI Detector}                                                              &  &  & First                                                      & Next                                                       & First                                                       & Next                                                          & \multirow{-2}{*}{FGSM}                                      & \multirow{-2}{*}{BIM}                                       \\ \hline
AmI Result                                                                                  &  &  & 0.97                                                       & 0.98                                                       & 0.85                                                        & 0.85                                                          & 0.91                                                        & 0.90                                                        \\ \hline
\begin{tabular}[c]{@{}c@{}}Our Result\\ (default $\beta$=60)\end{tabular}                   &  &  & 0.19                                                       & 0.40                                                       & 0.36                                                        & 0.44                                                          & 0.66                                                        & 0.61                                                        \\ \hline
{\color[HTML]{333333} \begin{tabular}[c]{@{}c@{}}Our Result\\ \textbf{(tuned $\beta$)}\end{tabular}} &  &  & \begin{tabular}[c]{@{}c@{}}0.97\\ ($\beta$=8)\end{tabular} & \begin{tabular}[c]{@{}c@{}}0.98\\ ($\beta$=8)\end{tabular} & \begin{tabular}[c]{@{}c@{}}0.85\\ ($\beta$=19)\end{tabular} & \begin{tabular}[c]{@{}c@{}}0.85\\ ($\beta$=16.5)\end{tabular} & \begin{tabular}[c]{@{}c@{}}0.91\\ ($\beta$=12)\end{tabular} & \begin{tabular}[c]{@{}c@{}}0.90\\ ($\beta$=14)\end{tabular} \\ \hline
\end{tabular}}
\end{table}

\textbf{The modified $\boldsymbol{\beta}$.} The absence of a clear definition for $\beta$ motivated us to adjust this parameter for our experimental analysis. To emphasize important features, it is essential to increase the values of correlated witness neurons. Based on the formula for $\beta$, it becomes clear that a smaller $\beta$ value is preferable for optimizing $\nu^{'}$. Consequently, we experimented with varying $\beta$ values in our subsequent tests. In particular, setting $\beta$ to 8 produced patch attack \cite{patch} accuracy results that were precisely aligned with those reported in \cite{AmI}.

\definecolor{CPPViolet} {HTML} {7040A0}
\lstset{
    columns=fixed,       
    frame=none,                                         
    backgroundcolor=\color[RGB]{245,245,244},           
    keywordstyle=\color[RGB]{40,40,255},                 
    language=python,                                        
    morekeywords={alignas,continute,friend,register,true,alignof,decltype,goto,
    reinterpret_cast,try,asm,defult,if,return,typedef,auto,delete,inline,short,
    typeid,bool,do,int,signed,typename,break,double,long,sizeof,union,case,
    dynamic_cast,mutable,static,unsigned,catch,else,namespace,static_assert,using,
    char,enum,new,static_cast,virtual,char16_t,char32_t,explict,noexcept,struct,
    void,export,nullptr,switch,volatile,class,extern,operator,template,wchar_t,
    const,false,private,this,while,constexpr,float,protected,thread_local,
    const_cast,for,public,throw,std},
    emph={map,set,multimap,multiset,unordered_map,unordered_set,
    unordered_multiset,unordered_multimap,vector,string,list,deque,
    array,stack,forwared_list,iostream,memory,shared_ptr,unique_ptr,
    random,bitset,ostream,istream,cout,cin,endl,move,default_random_engine,
    uniform_int_distribution,iterator,algorithm,functional,bing,numeric,},
    emphstyle=\color{CPPViolet}, 
}              
\begin{lstlisting}
def strengthen(x):
    return 2.15 - np.exp(-x/8)
\end{lstlisting} 

In light of this observation, we experimented with different values of $\beta$ in various attacks and succeeded in replicating the accuracy results reported in \cite{AmI} (as shown in Table \ref{results by tuning beta}). This outcome suggests that the high accuracy observed in \cite{AmI} results from adjusting $\beta$. This approach aligns with AmI's objective of emphasizing significant features by amplifying the discrepancy through a larger $\beta$. Inspired by these findings, we decided to concentrate on exploring this hyperparameter in our subsequent experiments.

\subsection{Adversarial Example Detection}

\cite{NC} asserts that \cite{AmI} does not resist the C\&W attack \cite{wb8}, as evidenced by experimental evaluation. The noted sensitivity to hyperparameters led us to employ a similar approach in assessing the C\&W attack \cite{wb8} as described in \cite{NC}. Although \cite{NC} does not provide the adversarial examples it generated nor the total count of these examples, it does open-source its codebase. This repository includes instructions for generating C\&W \cite{wb8} adversarial examples and the methodology for launching attacks against \cite{AmI}.

Using this available source code, we generated a total of 6705 C\&W \cite{wb8} adversarial examples. Our goal was to validate the claims made by \cite{NC} and further investigate \cite{AmI}'s ability to detect the C\&W attack \cite{wb8}.

The AmI \cite{AmI} is compatible with the Caffe and PyTorch frameworks. Our experiments were carried out using the Caffe framework.

\textbf{C\&W attack with the default $\boldsymbol{\beta}$.} In the code provided by \cite{NC}, the value of $\beta$ is set to the default of 60. Using this setting, we initiated a total of 6705 C\&W attacks \cite{wb8} against AmI \cite{AmI}. Of these, 4264 attacks were carried out successfully, with the distribution of results detailed in Table \ref{table:caffe_6705}.

\begin{table}[h!]
\centering
\setlength{\tabcolsep}{6mm}
\caption{Distribution of Cases using Caffe}
\begin{tabular}{|l|r|r|}
\hline
\textbf{Case} & \textbf{$\boldsymbol{\beta} = 60$} & \textbf{$\boldsymbol{\beta} = 5$} \\ 
\hline \hline
X - X - X & 4241 & 437 \\ 
\hline
X - X - Y & 10 & 4948 \\ 
\hline
X - Y - X & 4 & 0 \\ 
\hline
X - Y - Y & 3 & 0 \\  
\hline
X - Y - Z & 6 & 1320 \\  
\hline
\textbf{Total} & \textbf{4264} & \textbf{6705} \\  
\hline
\end{tabular}
\label{table:caffe_6705}
\end{table}

In scenarios classified as X - Y - Y, where the ground truth and the original model prediction differ, the example is identified as adversarial. However, because the original model's prediction aligns with AmI's \cite{AmI} prediction, AmI fails to recognize it as such, marking it as a false negative instance. This scenario is depicted in Figure \ref{fig:x-y-y}, where \cite{NC} specifically uses examples that fit this category to evaluate AmI. With $\beta$ set to 60, there are three instances of X - Y - Y cases, as shown in Table \ref{table:caffe_6705}. Using these cases to conduct C\&W attacks \cite{wb8}, the results, as shown in Table \ref{table:caffe_detection_rates}, reveal that with the default $\beta$, AmI is unable to detect the C\&W adversarial examples \cite{wb8}. This observation aligns with \cite{NC}'s assertion that the C\&W attack is effective against AmI.

\begin{table}[h!]
\centering
\caption{Comparison of AmI Detection Rates under Different Hyperparameters}
\begin{tabular}{|l|l|c|}
\hline
\textbf{Hyperparameter $\boldsymbol{\beta}$} & \textbf{Gold-Original-Attribute} & \textbf{Detection Rate} \\
\hline
\multirow{3}{*}{$\boldsymbol{\beta}=5$} & 202-1674-1484 & 1.00 [1/1] \\
 & 1997-32-31 & 1.00 [2/2] \\
 & 441-416-70 & 1.00 [3/3] \\
\hline
\multirow{3}{*}{Default $\boldsymbol{\beta}$} & 202-1674-1674 & 0.00 [0/1] \\
 & 1997-32-32 & 0.00 [0/2] \\
 & 441-416-416 & 0.00 [0/3] \\
\hline
\end{tabular}
\label{table:caffe_detection_rates}
\end{table}

\textbf{C\&W attack with the modified $\boldsymbol{\beta}$.} Having discovered the effectiveness of adjusting $\beta$, we hypothesize that \cite{NC} conducted their tests on AmI \cite{AmI} using its default $\beta$ value. To test our hypothesis, we reran the experiment with a modified $\beta$ value on 6705 C\&W \cite{wb8} adversarial examples. Remarkably, setting $\beta$ to 5 eliminated the occurrence of X - Y - Y cases, as evidenced in Table \ref{table:caffe_6705}. To further validate our findings, we applied AmI \cite{AmI} to analyze three X - Y - Y cases with $\beta$ adjusted to 5. The data presented in Table \ref{table:caffe_detection_rates} confirm that these three C\&W examples, previously misclassified at default $\beta$, are correctly identified with $\beta$ set to 5, allowing AmI to achieve the $100\%$ detection rate under the same experimental conditions established by \cite{NC}.

This outcome suggests that the outstanding detection accuracy of adversarial examples by AmI \cite{AmI} is the result of fine-tuning the hyperparameter $\beta$. Therefore, it would be prudent for AmI to document the variations in $\beta$ that produce significant accuracy improvements between different attacks and to suggest a systematic approach to adjust this parameter. However, such detailed guidelines or explanations regarding $\beta$ adjustments are absent in \cite{AmI}, to the best of our knowledge.

\subsection{Sensitivity to External Factors}
Interestingly, through our reproduction of \cite{AmI} and \cite{NC}, we found that AmI is also sensitive to factors that seem to be non-essential. Specifically, the performance of AmI \cite{AmI} is highly dependent on the operating system and deep learning framework used to conduct the experiment. For example, we present the detection rate of AmI \cite{AmI} after performing the attacks in Table \ref{table:combined_detection_rates}.

\begin{table}[h!]
\centering
\caption{Comparison of Detection Rates Across Environments}
\begin{tabular}{|l|c|c|}
\hline
\multicolumn{1}{|c|}{\textbf{Environment}} & \textbf{Gold-Original-Attribute} & \textbf{Detection Rate} \\
\hline
\multirow{6}{*}{Windows, PyTorch} & 202-202-2501 & 0.00 [0/0] \\
 & 441-441-717 & 0.00 [0/0] \\
 & 1997-32-845 & 1.00 [1/1] \\
 & 2230-2230-845 & 1.00 [1/1] \\
 & 2275-1550-1559 & 0.50 [1/2] \\
 & 2557-0-2557 & 0.67 [2/3] \\
\hline
\multirow{4}{*}{Linux, Caffe} & 202-18-18 & 0.00 [0/1] \\
 & 2557-0-0 & 0.00 [0/2] \\
 & 1997-32-32 & 0.00 [0/3] \\
 & 441-2282-2282 & 0.00 [0/4] \\
\hline
\multirow{4}{*}{Mac, PyTorch} & 202-18-18 & 0.00 [0/1] \\
 & 441-2282-2282 & 0.00 [0/2] \\
 & 1997-32-722 & 0.33 [1/3] \\
 & 2557-0-2557 & 0.50 [2/4] \\
\hline
\end{tabular}
\label{table:combined_detection_rates}
\end{table}


\textbf{Our conjecture}
We offer several insights regarding this observation. First, pre-trained VGG-Face PyTorch and Caffe models have different accuracy, which could in turn influence the efficacy of AmI's detection capabilities. Specifically, when presented with the same image, the VGG-Face model's outputs for \texttt{id\_original} may differ between the Caffe and PyTorch versions, due to discrepancies in accuracy. Moreover, AmI's methodology, which involves modulating neuron activation strengths, suggests that the underlying deep learning framework could be critical, given its role in defining neural network structures and performing computations. Similarly, the operating system can also affect performance. However, the ability to maintain consistent performance in various environments is crucial for the practical applicability of any defense technique.





\section{Conclusion and Future Work}
In this study, we explore the potential of using model explanation to detect adversarial examples. Building upon the evaluation of AmI by \cite{NC}, we discovered that while AmI may be effective, its robustness is compromised by variations in hyperparameter settings and external factors. We classify potential outcomes into five distinct cases, providing a detailed explanation for each. Moreover, we advocate for a more rigorous evaluation framework for adversarial example detection methodologies. Such a framework must adapt to environmental variations to mirror real-world applicability and consider both false positive and false negative rates.

We identify the following as promising avenues for future research:
\begin{itemize}
    \item Leverage model explanation to enhance our understanding of adversarial examples, particularly by figuring out the interpretable features that models rely on to make incorrect predictions.
    \item Employ model explanation techniques to develop robust methodologies to detect adversarial examples.
    \item Develop metrics and frameworks for a comprehensive evaluation of adversarial example detection methods.
\end{itemize}


\vspace{12pt}


\begin{thebibliography}{00}
\bibitem{b1} G. Eason, B. Noble, and I. N. Sneddon, ``On certain integrals of Lipschitz-Hankel type involving products of Bessel functions,'' Phil. Trans. Roy. Soc. London, vol. A247, pp. 529--551, April 1955.

\bibitem{AmI}
G.~Tao, S.~Ma, Y.~Liu, and X.~Zhang, ``Attacks meet interpretability: Attribute-steered detection of adversarial samples,'' 2018-10-27 2018.

\bibitem{NC}
N.~Carlini, ``Is ami (attacks meet interpretability) robust to adversarial examples?'' 2019.

\bibitem{Attribution-driven}
S.~Jha, S.~Raj, S.~L. Fernandes, S.~K. Jha, S.~Jha, G.~Verma, B.~Jalaian, and A.~Swami, ``Attribution-driven causal analysis for detection of adversarial examples,'' \emph{arXiv preprint arXiv:1903.05821}, 2019.

\bibitem{Interpretability}
J.~Wang, Y.~Wu, M.~Li, X.~Lin, J.~Wu, and C.~Li, ``Interpretability is a kind of safety: An interpreter-based ensemble for adversary defense,'' in \emph{Proceedings of the 26th ACM SIGKDD International Conference on Knowledge Discovery \& Data Mining}, 2020, pp. 15--24.

\bibitem{loo}
P.~Yang, J.~Chen, C.-J. Hsieh, J.-L. Wang, and M.~Jordan, ``Ml-loo: Detecting adversarial examples with feature attribution,'' in \emph{Proceedings of the AAAI Conference on Artificial Intelligence}, vol.~34, no.~04, 2020, pp. 6639--6647.

\bibitem{when}
G.~Fidel, R.~Bitton, and A.~Shabtai, ``When explainability meets adversarial learning: Detecting adversarial examples using shap signatures,'' in \emph{2020 international joint conference on neural networks (IJCNN)}.\hskip 1em plus 0.5em minus 0.4em\relax IEEE, 2020, pp. 1--8.

\bibitem{vgg}
W.~Xu, D.~Evans, and Y.~Qi, ``Feature squeezing: Detecting adversarial examples in deep neural networks,'' \emph{arXiv preprint arXiv:1704.01155}, 2017.

\bibitem{frs}
O.~M. Parkhi, A.~Vedaldi, and A.~Zisserman, ``Deep face recognition,'' 2015.

\bibitem{wb8}
N.~Carlini and D.~A. Wagner, ``Towards evaluating the robustness of neural networks,'' \emph{CoRR}, vol. abs/1608.04644, 2016. [Online]. Available: \url{http://arxiv.org/abs/1608.04644}

\bibitem{transferability}
N.~Papernot, P.~McDaniel, and I.~Goodfellow, ``Transferability in machine learning: from phenomena to black-box attacks using adversarial samples,'' \emph{arXiv preprint arXiv:1605.07277}, 2016.

\bibitem{carliniami}
N.~Carlini, ``Is ami (attacks meet interpretability) robust to adversarial examples?'' \emph{arXiv preprint arXiv:1902.02322}, 2019.

\bibitem{goodfellow2014explaining}
I.~J. Goodfellow, J.~Shlens, and C.~Szegedy, ``Explaining and harnessing adversarial examples,'' in \emph{International Conference on Learning Representations (ICLR)}, 2015.

\bibitem{moosavi2016deepfool}
S.-M. Moosavi-Dezfooli, A.~Fawzi, and P.~Frossard, ``Deepfool: a simple and accurate method to fool deep neural networks,'' in \emph{Proceedings of the IEEE conference on computer vision and pattern recognition}, 2016, pp. 2574--2582.

\bibitem{carlini2017towards}
N.~Carlini and D.~Wagner, ``Towards evaluating the robustness of neural networks,'' in \emph{2017 IEEE Symposium on Security and Privacy (SP)}.\hskip 1em plus 0.5em minus 0.4em\relax IEEE, 2017, pp. 39--57.

\bibitem{papernot2016limitations}
N.~Papernot, P.~McDaniel, I.~Goodfellow, S.~Jha, Z.~B. Celik, and A.~Swami, ``The limitations of deep learning in adversarial settings,'' in \emph{2016 IEEE European Symposium on Security and Privacy (EuroS\&P)}.\hskip 1em plus 0.5em minus 0.4em\relax IEEE, 2016, pp. 372--387.

\bibitem{kurakin2016adversarial}
A.~Kurakin, I.~Goodfellow, and S.~Bengio, ``Adversarial examples in the physical world,'' 2016.

\bibitem{liu2019transferable}
W.~Liu, Y.~Chen, Y.-W. Tai, and C.-K. Tang, ``Transferable adversarial attacks for image and video classification,'' 2019.

\bibitem{kurakin2016adversarialscale}
A.~Kurakin, I.~Goodfellow, and S.~Bengio, ``Adversarial machine learning at scale,'' 2016.

\bibitem{ilyas2018black}
A.~Ilyas, A.~Jalal, C.~Daskalakis, and A.~Madry, ``Black-box adversarial attacks with limited queries and information,'' in \emph{International Conference on Machine Learning}, 2018, pp. 2142--2151.

\bibitem{zhao2017adversarial}
S.~Zhao, Y.~Song, and S.~Ermon, ``Adversarial examples for generative models,'' 2017.

\bibitem{Szegedy2014Intriguing}
C.~Szegedy, W.~Zaremba, I.~Sutskever, J.~Bruna, D.~Erhan, I.~J. Goodfellow, and R.~Fergus, ``Intriguing properties of neural networks,'' in \emph{2nd International Conference on Learning Representations, {ICLR} 2014, Banff, AB, Canada, April 14-16, 2014, Conference Track Proceedings}, Y.~Bengio and Y.~LeCun, Eds., 2014. [Online]. Available:\url{http://arxiv.org/abs/1312.6199}

\bibitem{papernot2016distillation}
N.~Papernot, P.~McDaniel, X.~Wu, S.~Jha, and A.~Swami, ``Distillation as a defense to adversarial perturbations against deep neural networks,'' \emph{2016 IEEE Symposium on Security and Privacy (SP)}, pp. 582--597, 2016.

\bibitem{madry2017towards}
A.~Madry, A.~Makelov, L.~Schmidt, D.~Tsipras, and A.~Vladu, ``Towards deep learning models resistant to adversarial attacks,'' \emph{arXiv preprint arXiv:1706.06083}, 2017.

\bibitem{xu2018feature}
W.~Xu, D.~Evans, and Y.~Qi, ``Feature squeezing: Detecting adversarial examples in deep neural networks,'' in \emph{NDSS}, 2018.

\bibitem{cohen2019certified}
J.~Cohen, E.~Rosenfeld, and J.~Z. Kolter, ``Certified adversarial robustness via randomized smoothing,'' \emph{arXiv preprint arXiv:1902.02918}, 2019.

\bibitem{meng2017magnet}
D.~Meng and H.~Chen, ``Magnet: a two-pronged defense against adversarial examples,'' in \emph{Proceedings of the 2017 ACM SIGSAC Conference on Computer and Communications Security}.\hskip 1em plus 0.5em minus 0.4em\relax ACM, 2017, pp. 135--147.

\bibitem{song2017pixeldefend}
Y.~Song, R.~Shu, N.~Kushman, and S.~Ermon, ``Pixeldefend: Leveraging generative models to understand and defend against adversarial examples,'' in \emph{Proceedings of the International Conference on Learning Representations (ICLR)}, 2018.

\bibitem{lecuyer2019certified}
M.~Lecuyer, V.~Atlidakis, R.~Geambasu, D.~Hsu, and S.~Jana, ``Certified robustness to adversarial examples with differential privacy,'' in \emph{2019 IEEE Symposium on Security and Privacy (SP)}.\hskip 1em plus 0.5em minus 0.4em\relax IEEE, 2019, pp. 656--672.

\bibitem{samangouei2018defense}
P.~Samangouei, M.~Kabkab, and R.~Chellappa, ``Defense-gan: Protecting classifiers against adversarial attacks using generative models,'' in \emph{International Conference on Learning Representations}, 2018.

\bibitem{zheng2016improving}
S.~Zheng, Y.~Song, T.~Leung, and I.~Goodfellow, ``Improving the robustness of deep neural networks via stability training,'' in \emph{Proceedings of the IEEE Conference on Computer Vision and Pattern Recognition}, 2016, pp. 4480--4488.

\bibitem{wong2018provable}
E.~Wong, F.~R. Schmidt, J.~H. Metzen, and J.~Z. Kolter, ``Provable defenses against adversarial examples via the convex outer adversarial polytope,'' in \emph{Proceedings of the 35th International Conference on Machine Learning}, vol.~80, 2018, pp. 5283--5292.

\bibitem{patch}
T.~B. Brown, D.~Man{\'e}, A.~Roy, M.~Abadi, and J.~Gilmer, ``Adversarial patch,'' \emph{arXiv preprint arXiv:1712.09665}, 2017.

\bibitem{lime}
M.~T. Ribeiro, S.~Singh, and C.~Guestrin, ``Why should i trust you?: Explaining the predictions of any classifier,'' in \emph{Proceedings of the 22nd ACM SIGKDD International Conference on Knowledge Discovery and Data Mining}.\hskip 1em plus 0.5em minus 0.4em\relax ACM, 2016, pp. 1135--1144.

\bibitem{shap}
S.~M. Lundberg and S.-I. Lee, ``A unified approach to interpreting model predictions,'' in \emph{Advances in Neural Information Processing Systems}, vol.~30.\hskip 1em plus 0.5em minus 0.4em\relax Curran Associates, Inc., 2017.




\end{thebibliography}
\end{document}